# Bandwidth Adaptation for Scalable Videos over Wireless Networks

Mostafa Zaman Chowdhury[a,£], Tuan Nguyen[a], Young-Il Kim[b], Won Ryu[b], and Yeong Min Jang[a,*]
[a] Department of Electronics Engineering, Kookmin University, Korea
[b] Electronics and Telecommunications Research Institute (ETRI), Korea
E-mail: [£]mzceee@yahoo.com, [*]yjang@kookmin.ac.kr

*Abstract* —Multicast/broadcast services (MBS) are able to provide video services for many users simultaneously. Fixed amount of bandwidth allocation for all of the MBS videos is not effective in terms of bandwidth utilization, overall forced call termination probability, and handover call dropping probability. Therefore, variable bandwidth allocation for the MBS videos can efficiently improve the system performance. In this paper, we propose a bandwidth allocation scheme that efficiently allocates bandwidth among the MBS sessions and the non-MBS traffic calls (e.g., voice, unicast, internet, and other background traffic). The proposed scheme reduces the bandwidth allocation for the MBS sessions during the congested traffic condition only to accommodate more number of calls in the system. Our scheme allocates variable amount of bandwidths for the BMS sessions and the non-MBS traffic calls. The performance analyses show that the proposed bandwidth adaptation scheme maximizes the bandwidth utilization and significantly reduces the handover call dropping probability and overall forced call termination probability.

## I. Introduction

The existing wireless network technologies such as femtocell [1], WiFi, Mobile WiMAX, 3G, and 4G are able to support multicast/broadcast mechanisms [2], [3]-[6]. However, the bandwidth of existing wireless networks is still inadequate to support huge voice, data, and video services with full quality of service (QoS). To provide the high data rate video services e.g., multicast/broadcast services (MBS) and unicast services along with the existing voice, internet, and other background traffic services over the wireless networks, it is very important to efficiently manage the wireless bandwidth in order to ensure the admission of maximum number of calls in the system during the congested traffic condition, to maximize the overall service quality, and to maximize the review.

Scalable video technique [2], [7]-[9] allows the variable bit rate video broadcast/multicast/unicast over wireless networks. This technique utilizes multiple layering. Each of the layers improves spatial, temporal, or visual quality of the rendered video to the user [2]. Base layer or the highest priority layer guarantees the minimum quality of a video stream. The addition of any enhanced layer or low priority layer improves the video quality. The number of layers for a video session and the bandwidth per layer can be manipulated dynamically. Therefore, to broadcast/multicast/unicast videos through a wireless network, layered transmission is an effective approach for supporting heterogeneous receivers with varying bandwidth requirements [9].

In this paper, we propose a bandwidth adaptation based bandwidth allocation scheme that efficiently allocates bandwidth among the MBS sessions and the non-MBS traffic calls (e.g., voice, unicast video, internet, and other background traffic). The proposed scheme decreases the bandwidth allocation for each of the MBS sessions during the congested traffic condition only to accommodate more calls in the system. Our scheme allocates variable amount of bandwidths for them. However, the minimum quality of each of the videos is guaranteed by allocating minimum bandwidth for each of the video sessions. The SVC technique allows the reduced bandwidth allocation for the MBS sessions and the unicast videos. The proposed scheme also reduces the bandwidth allocation for the background traffic based on the QoS adaptability [10], [11] of the multimedia traffic.

The rest of this paper is organized as follows. Section II presents the proposed bandwidth adaptation scheme. Performance evaluation results of the proposed scheme are presented and compared with other schemes in Section III. Finally, Section IV concludes our work.

## II. Proposed Bandwidth Adaptation Scheme

Fig. 1 shows the basic concept of bandwidth allocations for proposed scheme.

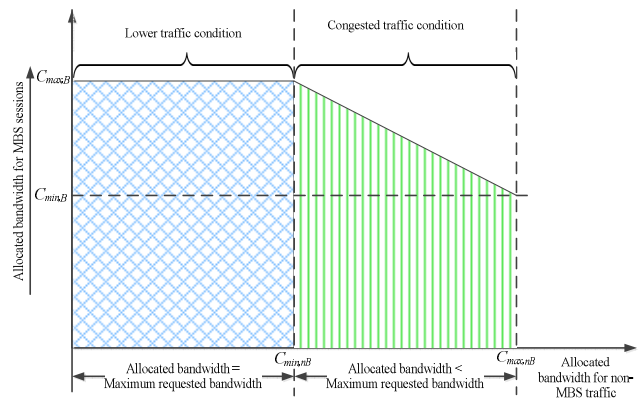

**Fig. 1.** Bandwidth allocation concept to allocate bandwidth among the MBS sessions and the non-MBS traffic calls.

---

[*]Corresponding author of the paper

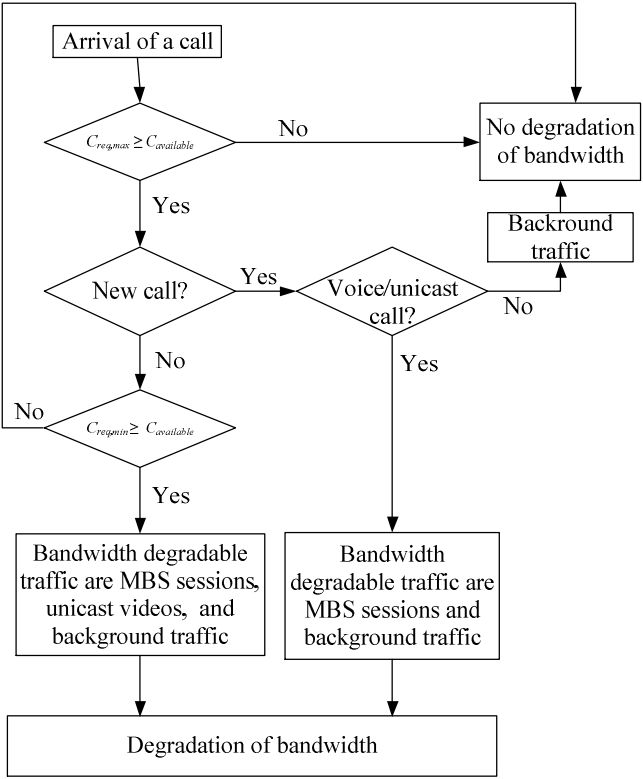

**Fig. 2.** Proposed bandwidth degradation policy.

For the low traffic condition, all of the calls are provided with the maximum qualities. However, for the congested traffic condition, the bandwidth allocation for the MBS sessions and the non-MBS traffic calls are decreased. Suppose $C_{max,nB}$ and $C_{min,nB}$ are, respectively, the maximum allowable and the minimum allowable bandwidths for the non-MBS traffic calls. $C_{max,B}$ and $C_{min,B}$ are, respectively, the maximum allowable and the minimum allocated bandwidths for the active MBS video sessions. The bandwidth $C_{max,B}$ is provided to MBS sessions only if the allocated bandwidth for the non-MBS traffic calls is less than or equal to $C_{min,nB}$.

The proposed scheme permits reclaiming some of the bandwidth from already admitted QoS adaptive multimedia traffic calls and MBS sessions, as to accept more calls in the system. Therefore, our scheme can accommodate more calls. Fig. 2 shows the procedure of bandwidth degradation for the proposed scheme. The proposed scheme gives highest priority for the handover calls. Suppose $C_{req,max}$ and $C_{req,min}$ are, respectively, the maximum and the minimum required bandwidths for a requested call. The system accepts a handover call if it can manage only $C_{req,min}$ amount of bandwidth. However, for the case of new call arrival, it is equal to $C_{req,max}$. The qualities of the unicast video calls are degraded only to accept handover calls in the system. The overall resource allocation scheme is divided into four categories based on the traffic characteristics. The resource allocation and QoS adaptation for each of the traffic types is different.

Suppose $\beta_v$, $\beta_{min,v}$, and $\beta_{max,v}$ are, respectively, the currently allocated, minimum allocated, and maximum allocated bandwidths for each of the voice calls. The bandwidth relation for the voice calls is found as:

$$\beta_v = \beta_{min,v} = \beta_{max,v} \quad (1)$$

The bandwidth relationships for the unicast video calls are stated as follows:

$$\beta_{min,uni} = \beta_{0,uni} + \beta_{1,uni} + \cdots + \beta_{K_{min},uni} \quad (2)$$

$$\beta_{max,uni} = \beta_{0,uni} + \beta_{1,uni} + \cdots + \beta_{K_{min},uni} + \cdots + \beta_{K_{max},uni} \quad (3)$$

$$\beta_{uni} = \beta_{0,uni} + \beta_{1,uni} + \cdots + \beta_{K_{min},uni} + \cdots + \beta_{k,uni} \quad (4)$$

where $\beta_{uni}$, $\beta_{min,uni}$, and $\beta_{max,uni}$ are, respectively, the currently allocated bandwidth, minimum allocated bandwidth, and maximum allocated bandwidth for each of the unicast video calls. $\beta_{0,uni}$ is the allocated bandwidth for the base layer of each of the unicast video calls. $K_{max}$ and $K_{min}$ are, respectively, the maximum and the minimum numbers of supported enhanced layers for each of the unicast video calls. $\beta_{k,uni}$ is the required bandwidth for the $k$-$th$ layer of a unicast call.

The bandwidth relationships for the MBS video sessions are expressed as follows:

$$\beta_{min,m} = \beta_{0,m} + \beta_{1,m} + \cdots + \beta_{N_{min,m},m} \quad (5)$$

$$\beta_{max,m} = \beta_{0,m} + \beta_{1,m} + \cdots + \beta_{N_{min,m},m} + \cdots + \beta_{N_{max,m},m} \quad (6)$$

$$\beta_{B,m} = \beta_{0,m} + \beta_{1,m} + \cdots + \beta_{N_{min,m},m} + \cdots + \beta_{N_m,m} \quad (7)$$

where $\beta_{B,m}$, $\beta_{min,m}$, and $\beta_{max,m}$ are, respectively, the currently allocated, minimum allocated, and maximum allocated bandwidths for $m$-$th$ MBS session. $\beta_{0,m}$ is the allocated bandwidth for the base layer of the $m$-$th$ MBS session. $N_{max,m}$ and $N_{min,min}$ are, respectively, the maximum and the minimum numbers of supported enhanced layers for the $m$-$th$ MBS session. $\beta_{n,m}$ is the required bandwidth for the $n$-$th$ layer of the $m$-$th$ MBS session. $M$ is the number of active MBS sessions.

If $C_{nB}$ is the allocated bandwidth for the non-MBS traffic calls, then the lower traffic condition is defined as $C - C_{nB} \geq \sum_{m=1}^{M} \sum_{n=0}^{N_{max,m}} \beta_{n,m}$. For this condition, the allocated bandwidth for the non-MBS traffic calls is less than or equal to the $C_{min,nB}$. Therefore, all the MBS sessions are provided with the maximum allowable bandwidth. The allocated bandwidth ($C_B$) for the MBS sessions during the lower traffic condition is calculated as,

$$C_B = C_{max,B} = \sum_{m=1}^{M} \sum_{n=0}^{N_{max,m}} \beta_{n,m}$$
$$\beta_{B,m} = \beta_{max,m} = \beta_{0,m} + \beta_{1,m} + \cdots + \beta_{N_{min,m},m} + \cdots + \beta_{N_{max,m},m} \quad (8)$$

Congested traffic condition is defined as, $C - C_{nB} < \sum_{m=1}^{M} \sum_{n=0}^{N_{max,m}} \beta_{n,m}$. For this congested traffic condition, the assigned bandwidth for the non-MBS traffic calls is greater than $C_{min,nB}$. Therefore, all the MBS sessions are not provided with the maximum allowable bandwidth.

The proposed scheme provides almost equal degradation of MBS video qualities. The maximum difference between the reduced number of enhanced layers for two MBS sessions is one. Therefore, if the reduced number of enhanced layers for the most popular video session is $P$, the reduced number of enhanced layers for the lowest popular video session is either $P$ or $(P+1)$. The bandwidth for each of the MBS sessions is calculated as follows:

If $C - C_{nB} < \sum_{m=1}^{M} \sum_{n=0}^{N_{max,m}} \beta_{n,m}$,

$$\beta_{B,m} = \begin{cases} \beta_{0,m} + \beta_{1,m} + \cdots + \beta_{N_{min,m},m} + \cdots + \beta_{N_{max,m}-P,m}, & 1 \le m \le M_1 \\ \beta_{0,m} + \beta_{1,m} + \cdots + \beta_{N_{min,m},m} + \cdots + \beta_{N_{max,m}-P-1,m}, & M_1 < m \le M \end{cases}$$
(9)

$$C_B = \sum_{m=1}^{M_1} \sum_{n=0}^{N_{max,m}-P} \beta_{n,m} + \sum_{m=M_1+1}^{M} \sum_{n=0}^{N_{max,m}-P-1} \beta_{n,m} \quad (10)$$

where $P$ is the minimum number of enhanced layers that is removed from every active MBS sessions due to the congested traffic condition. $M_1$ is the minimum number of MBS sessions for which $P$ number of enhanced layers are removed and for the remaining $(M-M_1)$ number of MBS sessions $(P+1)$ number of enhanced layers are removed.

The bandwidth relations for the background traffic are expressed as follows:

$$\beta_{min,back(i)} = (1 - \xi_i) \beta_{max,back(i)} \quad (11)$$

where $\beta_{min,back(i)}$ and $\beta_{min,back(i)}$ are, respectively, the minimum and maximum allocated bandwidths for a background traffic call of $i$-th class. $\xi_i$ is the maximum levels of bandwidths that can be degraded for a background traffic call of $i$-th class.

## III. Performance Analysis

In this section, we present the results of the numerical analysis of the proposed scheme. We compared the performance of our proposed scheme with the performance of the "fixed bandwidth allocation for MBS sessions" schemes. Table 1 shows the assumptions of the summary of the parameter values used in our analysis. The call arriving process and the cell dwell times are assumed to be Poisson. The average cell dwell time is assumed to be 540 sec [12].

Fig. 3 shows that the proposed scheme provides negligible handover call dropping probability even for very high traffic condition. The 14 Mbps for MBS sessions causes the reduced maximum bandwidth allocation for the non-MBS traffic calls. This reduced bandwidth allocation for the non-MBS traffic calls and also the non-priority of handover calls causes very high handover call dropping probability. 6 Mbps for MBS sessions provides higher bandwidth for the non-MBS traffic calls but the non-priority of handover calls causes high handover call dropping probability.

**Table 1.** Summary of the parameter used in analysis

| Parameter | Value |
| --- | --- |
| Bandwidth capacity ($C$) | 20 Mbps |
| Required/allocated bandwidth for each of the voice calls ($\beta_v$) | 64 kbps |
| Maximum allocated bandwidth for each of the unicast video calls ($\beta_{max,uni}$) | 0.5 Mbps |
| Minimum allocated bandwidth for each of the unicast video calls ($\beta_{min,uni}$) | 0.3 Mbps |
| Maximum allocated bandwidth for each of the MBS sessions ($\beta_{max,m}$) | 1 Mbps |
| Minimum allocated bandwidth for each of the MBS video sessions ($\beta_{min,m}$) | 0.5 Mbps |
| Number of MBS sessions ($M$) | 12 |
| Maximum required/allocated bandwidth for each of the background traffic calls ($\beta_{max,back}$) | 120 kbps |
| Minimum required/allocated bandwidth for each of the background traffic calls ($\beta_{min,back}$) | 60 kbps |
| Ratio of call arrival rates (voice: unicast call: background traffic ) | 5:1:4 |

Fig. 4 shows the overall forced call termination probability performance comparison. Our proposed scheme provides best performance due to the dynamic nature of bandwidth allocation both for the MBS sessions and the non-MBS traffic calls. The other two schemes cannot improve the overall forced call termination performance due to the QoS non-adaptive nature of both the schemes and also the reduced allocated bandwidth for the scheme where 14 Mbps is allocated for the MBS sessions.

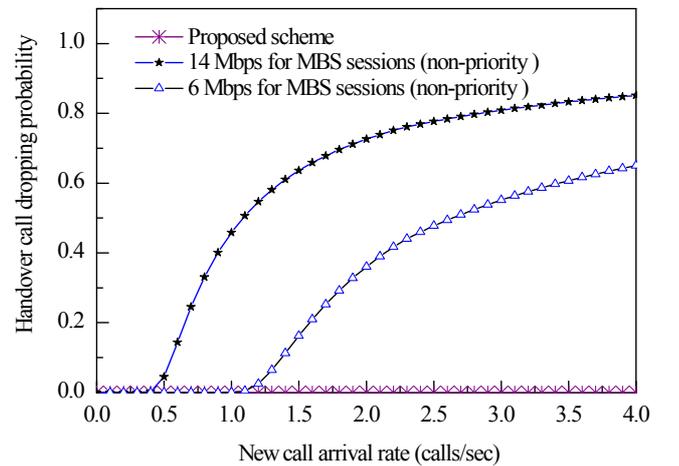

**Fig. 3.** Comparison of handover call dropping probability.

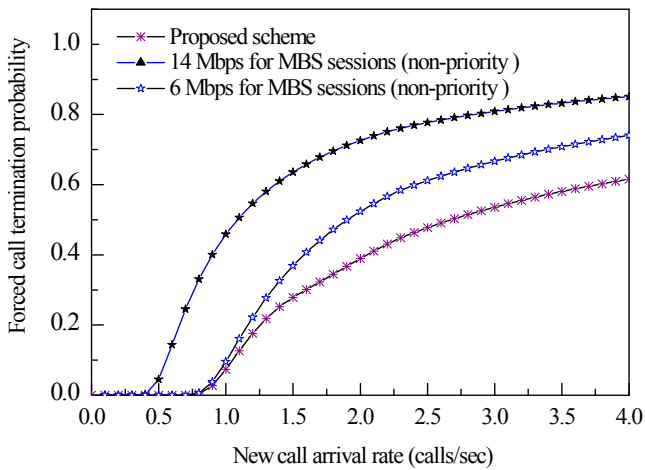

**Fig. 4.** Comparison of overall forced call termination probability

## IV. Conclusions

Video over the wireless link is one of the most promising services for the current and next generation communication. In this paper, we proposed a QoS adaptive bandwidth allocation scheme for MBS videos over wireless networks. Our idea behind the proposed scheme is that, during the congested traffic condition, the system releases some part of bandwidth from the MBS video sessions and other running QoS adaptive calls to accommodate more number of calls in the system. More bandwidth is released to support handover calls over new calls. Also, more bandwidth is released to support new voice and unicast video calls over new background traffic calls. Thus, the scheme results in negligible handover call dropping probability for all traffic types and lower new call blocking blocking probability for voice call and unicast calls.

The proposed scheme provides the opportunity for the network operator to increase the revenue. Therefore, the proposed scheme is expected to be of considerable interest for MBS provision through wireless networks.

## Acknowledgments

This work was supported by Electronics and Telecommunications Research Institute (ETRI), Korea